# Inverse Low-Dimensional Manifold Reconstruction Framework for Spatiotemporal Reconstruction of Compressible Physical Fields

Qiang Liu,[a)], Feng Ma,[a)], Wei Zhu,[a,*)], Xiyu Jia,[a)], Jianmin Xue,[a)] ,Jun Wen,[a)], Gaojun Fu,[a)]

**AFFILIATIONS**

[a] Beijing Institute of Technology, Beijing 100081, China

* Correspondence author:

zhu.wei@bit.edu.cn(Wei Zhu)

## Abstract

Compressible physical fields are widely present in the real physical world, but current artificial intelligence lacks an understanding mechanism for the non-differentiable features in compressible physical fields. Addressing the limitations of existing deep learning architectures in handling global non-differentiable features, we propose the Inverse Low-Dimensional Manifold reconstruction framework (ILDM). This framework couples the Non-differentiable Approximation Function (NAF) for capturing non-differentiable features in compressible flows with the Smooth Fluid Reconstruction (SFR) module tailored for smooth fluid regions. Extensive evaluations across 1D and 2D benchmarks, including Riemann problems and double Mach reflection, demonstrate that ILDM significantly outperforms cPINN and R-adaptive DeepONet. Specifically, ILDM achieves superior localization of non-differentiable interfaces and maintains robust super-resolution performance even with low-resolution inputs, establishing a physically consistent and scalable paradigm for data-driven fluid dynamics.

## Introduction

The occurrence of non-differentiable phenomena in physical variables is an inherent and unavoidable challenge in the study of compressible flows[1]. The reconstruction of such non-differentiable features provides a fundamental basis for the development of high-order numerical algorithms and physically consistent data-driven models. The governing equations of compressible flows typically admit non-smooth solutions; driven by the convergence of characteristic lines and the associated multi-valued paradoxes, compressible flow fields manifest a pronounced structural heterogeneity, characterized by the coexistence of non-differentiable interfaces and locally smooth regions[2-4]. This heterogeneity poses significant challenges for the high-fidelity reconstruction of physical fields. Traditional numerical methods—relying on the integral form of conservation laws, numerical flux functions, and Courant–Friedrichs–Lewy (CFL) stability constraints—remain susceptible to Gibbs phenomena in the presence of discontinuities, often struggling to maintain a balance between high-order accuracy and numerical robustness. Recently, the advent of deep learning has introduced hybrid computational strategies that integrate neural networks with physical priors[5]. These approaches, operating under supervised or self-supervised paradigms, enable the learning and prediction of physical fields from data, offering a promising alternative to mitigate the inherent limitations of conventional numerical schemes.

In recent years, a broad spectrum of neural network architectures has been explored for solving non-differentiable physical dynamics[6-9]. For forward and inverse problems in continuum physics, Karniadakis and co-workers pioneered the Physics-Informed Neural Network (PINN) framework[10]. However, the presence of non-differentiable interfaces and derivative singularities in non-differentiable fields severely limits the applicability of standard PINNs. To address this issue, Karniadakis subsequently introduced the conservative PINN (cPINN)[11], which decomposes the computational domain into multiple subdomains, trains an independent PINN in each subdomain, and enforces Rankine–Hugoniot jump conditions at the interfaces between subdomains. Nonetheless, the locations of these subdomain interfaces still need to be prescribed manually. Moreover, other researchers have developed computational models for non-differentiable physical dynamics based on the PINN architecture[12-15], but these are, in essence, closely related to and highly similar to cPINN. Deep Operator Network (DeepONet)[16], proposed by Lu et al., encounters similar difficulties when applied to non-differentiable problems, as DeepONet is essentially a smooth operator approximator and cannot inherently capture jump-type behaviour. To address this issue, Ahmad Peyvan et al. employed DeepONet to directly learn the solution procedure of Riemann problems[17], enforcing the Rankine–Hugoniot conditions on the propagation speed of discontinuities, with the left and right states as inputs and the resulting wave structure as the output. Yameng Zhu et al. further proposed an R-adaptive DeepONet that repeatedly redistributes training samples so that a larger fraction of points is concentrated near non-differentiable regions[18]. PINN and DeepONet-based strategies can achieve good performance for relatively simple, low-dimensional discontinuities[19,20].

Reconstructing the physical dynamics of compressible flow fields fundamentally necessitates deciphering their underlying spatiotemporal reconstruction mechanisms[21,22]. However, the global spatial reconstruction of such fields remains beyond the current representational limits of deep learning models. To circumvent this challenge, existing architectures often employ non-differentiable interfaces as computational boundaries to decompose the spatial domain[23-25]. While this strategy is designed to preserve numerical accuracy, it implicitly necessitates that these interfaces remain stationary or move at negligible velocities. Such an assumption stands in stark contrast to the physical reality of compressible flows, where the rapid propagation of non-differentiable interfaces—such as shock waves—is ubiquitous.

To achieve global reconstruction of compressible flow fields, we explicitly construct non-differentiable interfaces by leveraging non-differentiable gradient structures and manifold representations. In compressible physical fields, the divergence and confluence of characteristic lines give rise to expansion waves and shock waves, which manifest as continuous yet non-smooth transitions and discrete discontinuities, respectively[26-29]. In 2D and 3D flow fields, the interaction of these diverse wave topologies results in intricate wave systems and induces transitions in the underlying flow states between individual waves[30]. To reconstruct such complex non-differentiable interfaces, we adopt a manifold-based perspective to simplify the interfaces through dimensionality compression. Subsequently, high-dimensional global non-differentiable features are recovered from these low-dimensional representations via a dimensionality lifting process.

Consequently, we develop a computational framework (ILDM) tailored for the global reconstruction of compressible physical fields, comprising two core components: a non-differentiable approximation function, and a smooth fluid reconstruction module. Designed for universal applicability across diverse compressible flow scenarios, the proposed model is characterized by three primary features:

1. **Explicit Non-differentiable Approximation**: We introduce an explicitly represented NAF. By constructing bidirectional Sigmoid functions integrated with internal scaling factors, the model can effectively approximate non-differentiable features. Simultaneously, the spatiotemporal evolution of non-differentiable interfaces is captured through the learning capabilities of neural networks.
2. **Spatiotemporal Reconstruction of Smooth Flow**: To address the smooth flow regions interspersed between non-differentiable interfaces, we leverage a Convolutional LSTM (ConvLSTM) architecture. By integrating convolutional operators with temporal recurrent units, the model effectively learns the spatiotemporal dynamics of smooth fluid regions.
3. **Inverse Low-Dimensional Manifold reconstruction framework**: Recognizing that compressible flows in general scenarios typically reside in spatial domains of two or more dimensions, we employ a manifold-inspired approach to perform dimensionality lifting of non-differentiable features. Specifically, these features undergo a progressive reconstruction from zero-dimensional points to one-dimensional curves, subsequently expanding into non-differentiable interfaces. Ultimately, the global reconstruction of the entire compressible flow field is achieved by seamlessly integrating these interfaces with the smooth flow regions.

In this study, the efficacy and robustness of the resulting ILDM are demonstrated through its application to a wide range of compressible physical field reconstruction tasks.

**Workflow**

Compressible physical fields are principally composed of complex non-differentiable interfaces and smooth fluid regions. Both the non-differentiable interfaces and the smooth flow structures represent dynamic fields that undergo spatiotemporal evolution under prescribed initial and boundary conditions. We formulate the compressible physical field as a linear superposition of these two components, expressed as follows:

$$U_0 : D_0 + C(x_C, t_0),$$
$$x_C \in \Omega_C = \{x \mid \sigma'(C(x,t_0)) \neq 0\}$$

where $x$ denotes the spatial variable, $t_0$ is the initial time, $U_0$ is the physical field at the initial time, $D_0$ is the non-differentiable interface, $C$ is the smooth fluid structures within the physical field, and $\Omega_C$ is the smooth-flow subdomain.

Consequently, the reconstruction of compressible flow fields is decomposed into two distinct stages. First, a non-differentiable approximation function is designed based on the gradient characteristics of the interfaces. This function must concurrently approximate two prevalent types of interfaces in compressible flows: discontinuous and continuous yet non-differentiable. Specifically, the characterization of non-differentiable requires the function's derivative at the singular point to significantly exceed that of its neighboring points. To satisfy this requirement, we formulate the fundamental structure of the approximation function as the product of a monotonically increasing and a monotonically decreasing Sigmoid function. Building upon this structural baseline, a gradient scaling factor and a peak coefficient are incorporated, rendering the gradient variation trends on both sides of the non-differentiable point tunable:

$$d = F_{\varphi_d,\phi_d}(x; v^i, x^i) = \sum_{i=1}^{n} v^i * \frac{1}{1+e^{(x-x^i)*\varphi_d}} * \frac{1}{1+e^{-(x-x^i)*\phi_d}}$$

where $d$ denotes the non-differentiable curves, $F_{\varphi_D,\phi_D}$ is the non-differentiable approximation function, $n$ is number of non-differentiable points, $v^i$ is a numerical scaling coefficient, $x^i$ is the location of the non-differentiable points, $\varphi_d$ and $\phi_d$ are the forward and backward scale factors, respectively; the design rationale of this function is detailed in the Methods section.

Conversely, the approximation of continuous non-differentiable transitions requires the derivative at the singular point to significantly exceed that of its neighbors on only one side. This specific profile can be synthesized through the superposition of non-differentiable approximation functions characterized by disparate gradient trends, as illustrated in Fig. 1b.

Through this construction methodology, we leverage a suite of parametrically controllable functions to reconstruct the intrinsic non-differentiable features of flow interfaces. Building upon this functional basis, we developed the ILDM to recover the spatial architecture of these interfaces. Given that non-differentiable interfaces can be conceptualized as low-dimensional manifolds embedded within compressible physical fields—where non-differentiable points and curves represent even lower-dimensional manifold substructures—we propose an inverse dimensionality reconstruction strategy characterized by dimensionality lifting. Specifically, the procedure involves the nested application of non-differentiable approximation functions, progressively achieving global reconstruction of the interface by evolving from zero-dimensional points to one-dimensional curves and ultimately to two-dimensional surfaces.

$$f_k = \sum_{i=1}^{n} F_{\varphi_d,\phi_d}(x; v_k^i, x_k^i) + c(x_c, t_k) \ , \quad x_c \in \Omega_c$$

where $f_k$ denotes the non-differentiable curve, $v_k^i$ is the height difference between non-differentiable points, $x_k^i$ is the non-differentiable points, $\varphi_d$ is the forward scale factor, $\phi_d$ is the backward scale factor, and $c(x_c, t_k)$ is the smooth fluid structures along the curve, $\Omega_c$ is the smooth region along the non-differentiable curve. The complete time-evolving compressible physical field is then expressed as:

$$\begin{aligned} &U_k : D_k + C(x_C, t_k), \\ &D_k = \sum_{j=1}^{m} F_{\varphi_D,\phi_D}(x; v_k^j, f_k^j) \\ &f_k = \sum_{i=1}^{n} F_{\varphi_d,\phi_d}(x; v_k^i, x_k^i) + c(x_c, t_k), \\ &x_C \in \Omega_C, x_c \in \Omega_c \end{aligned}$$

where $U_k$ denotes the non-differentiable physical field at time $t_k$, and $t_k$ is the $k$-th time instant. During computation, $v_k^j$ is determined by the initial conditions. $x_k^i$ 、 $v_k^i$ 、 $C(x_C, t_k)$ and $c(x_c, t_k)$ are learned by different neural networks, and the complete computational pipeline for the time-evolving compressible physical field is illustrated in Fig. 1b.

Subsequently, the reconstruction of smooth flow structures necessitates masking the compressible physical field, utilizing the previously identified non-differentiable interfaces as a spatial template. This procedure effectively isolates the smooth regions and circumvents the detrimental impact of non-differentiability on the reconstruction accuracy. The spatiotemporal reconstruction of smooth flow requires the simultaneous extraction of spatial morphological features and temporal evolutionary dynamics. To this end, we develop a smooth-flow reconstruction module based on the ConvLSTM architecture. Within this framework, convolutional operators are employed to extract high-dimensional spatial features, while the gating mechanisms of the LSTM units are leveraged to capture and preserve the underlying spatiotemporal dynamics of the smooth fluid.

The global reconstruction of the compressible physical field is finalized through the spatial superposition of the smooth fluid regions and the non-differentiable interfaces. The comprehensive workflow of this integration process is schematically illustrated in Fig. 1.

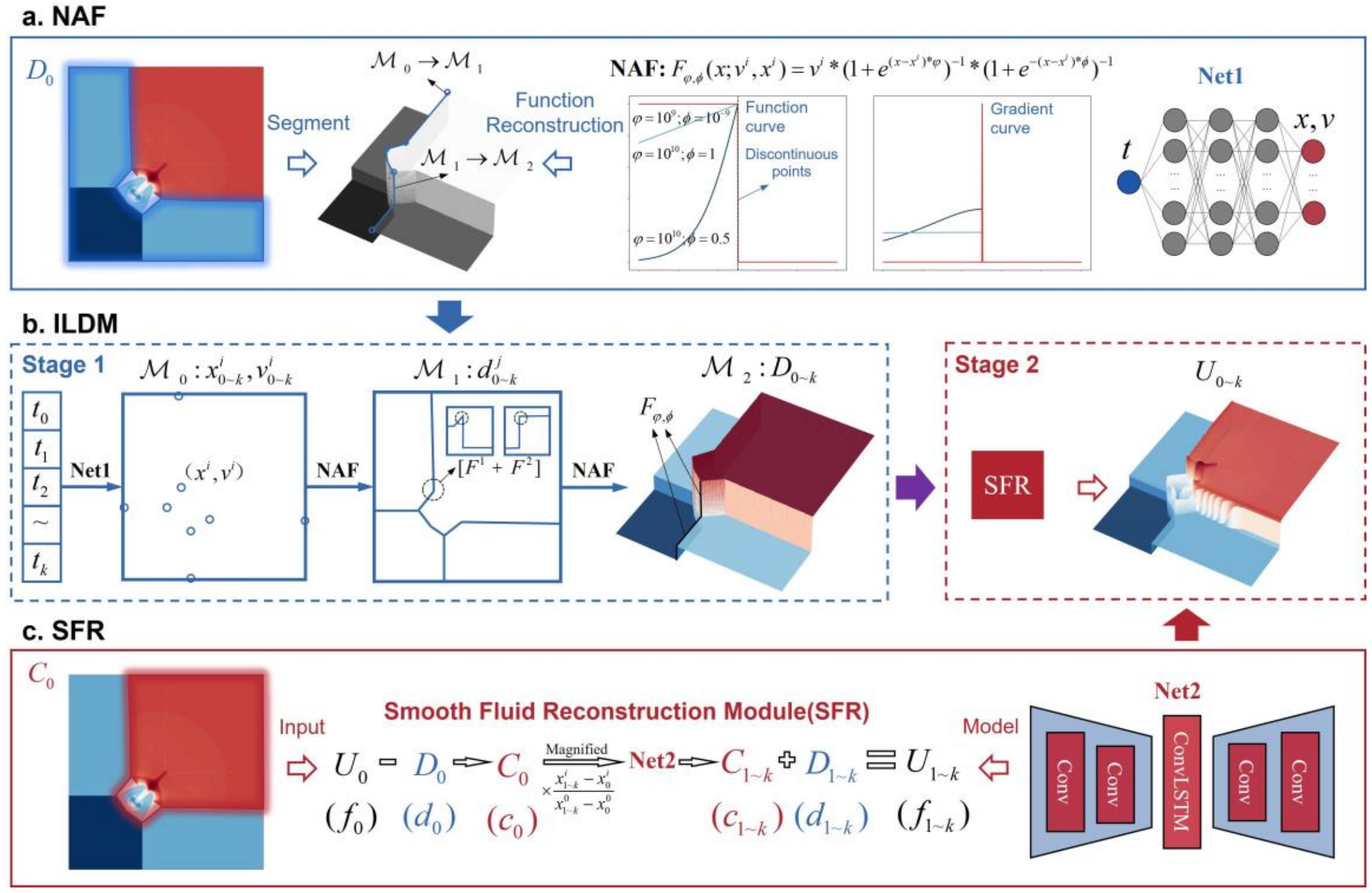

**Fig. 1 | Scheme for reconstructing compressible physical fields. a**, Extraction of non-differentiable surfaces from a non-differentiable physical field and application of the

NAF to reconstruct these non-differentiable surfaces. **b**, Inverse Low-Dimensional Manifold reconstruction framework. The NAF uses the propagation behaviour of non-differentiable points $x^i_{0\sim k}, v^i_{0\sim k}$ to reconstruct the non-differentiable structure $d^j_{0\sim k}$ along a non-differentiable curve; the non-differentiable curve $f^j_0$ is then reconstructed by combining $d^j_{0\sim k}$ and $c(x_c, t_k)$, and the same procedure is iterated to obtain the full time-evolving non-differentiable physical field. **c,** Smooth fluid reconstruction module. The SFR module takes the smooth fluid regions within the compressible flow field as its input, utilizing a ConvLSTM network as the core reconstruction model.

## Methods

The Inverse-Manifold Network proposed in this work consists of three main components: the NAF, the ILDM architecture and the SFR. In the following, we describe the use of NAF, ILDM and SFR in detail.

*NAF*

The primary objective of the NAF ($F_{\varphi,\phi}(x; v^i, x^i)$) is to reconstruct the gradient characteristics of non-differentiable interfaces within compressible flow fields, while providing functional support for the dimensionality reconstruction executed by the ILDM. Structurally, the $F_{\varphi,\phi}(x; v^i, x^i)$ is composed of two Sigmoid functions with opposing variation trends. Its explicit parameters include a gradient scaling factor $\varphi, \phi$, a peak coefficient $v^i$, and a positional variable $x^i$. The adjustment of the gradient scaling factor is specifically tailored to the morphological characteristics of the non-differentiable interface. In typical configurations, the gradient distribution near such interfaces manifests in two characteristic modes: (i) the gradient at the non-differentiable point significantly exceeds those of its neighbors, and (ii) the gradient at the non-differentiable point exceeds those of its neighbors on only one side. The first mode is realized by modulating the gradient scaling factor within the NAF to induce a steep gradient ascent at the interface. In contrast, the second mode necessitates the simultaneous adjustment of bidirectional scaling factors and the superposition of two NAFs with opposing orientations to achieve a non-zero gradient on a single side. The mapping for the iterative updating of $v^i$ and $x^i$ from the temporal variable $t_k$ is implemented via a Multi-Layer Perceptron (MLP), which is constructed by stacking multiple fully connected layers:

$$v^i(t) := \sigma(\mathbf{W}^L\sigma(\mathbf{W}^{L-1},\cdots,\sigma(\mathbf{W}^1 t+\mathbf{b}^1)+\mathbf{b}^{L-1})+\mathbf{b}^L),$$
$$x^i(t) := \sigma(\mathbf{W}^L\sigma(\mathbf{W}^{L-1},\cdots,\sigma(\mathbf{W}^1 t+\mathbf{b}^1)+\mathbf{b}^{L-1})+\mathbf{b}^L),$$

The learning process of the height and position variables by the MLP can be expressed mathematically as:

$$\theta_{v^i} \in \arg\min \frac{1}{N}\sum_{i=1}^{N}\left\| v^i - \hat{v}(t_i)\right\|_2,$$

$$\theta_{x^i} \in \arg\min \frac{1}{N}\sum_{i=1}^{N}\left\| x^i - \hat{x}(t_i)\right\|_2.$$

*SFR*

NAF and ILDM are primarily designed for computing non-differentiable interfaces in non-differentiable physical dynamics, whereas a wide range of existing models can be used to reconstruct the smooth fluid structures. On top of ILDM, we therefore develop a SFR that embeds reliable continuous-flow reconstruction models (such as ConvLSTM and U-Net) into the framework. Within SFR, we first use the initial non-differentiable flow field $U_0$ together with the non-differentiable interface $D_0$ to separate the smooth fluid structures $C_0$. We then treat the non-differentiable interface as a boundary and employ an existing reconstruction model to reconstruct the smooth fluid structures $C_k$ at time $t_k$. Finally, by combining NAF and ILDM, we obtain the non-differentiable interfaces $D_k$ and $C_k$ at time $t_k$, and reconstruct the corresponding non-differentiable flow field $U_k$. The loss function used to train the continuous reconstruction model is given mathematically by:

$$\theta_{C_k} \in \arg\min \frac{1}{N}\sum_{i=1}^{N}\left\| C_k - \hat{C}(t_i)\right\|_2.$$

where $\hat{C}$ denotes the smooth fluid reconstruction model. The reconstruction procedure for the smooth fluid structures $c_k$ along the non-differentiable interface is identical to that used for $C_k$.

*ILDM*

Building upon the NAF, we construct an ILDM architecture designed to recover the spatial characteristics of non-differentiable interfaces. Each interface is conceptualized as an $(n-1)$-dimensional Lipschitz manifold embedded within the compressible physical field $\Omega \subset \mathbb{R}^n$, while the non-differentiable curves and points situated on the interface constitute its 1D and 0D Lipschitz submanifolds, respectively. To reconstruct these geometric structures, we leverage a level-set mapping $F_{\varphi,\phi}(x; v^i, x^i) \in \mathrm{Lip}(\Omega)$ induced by the NAF and define a Lipschitz-class embedding $\Phi_k : \mathcal{M}_k \to \mathcal{M}_{k+1}$ that maps $k$-dimensional structures to $(k+1)$-dimensional interfaces. This constructive approach facilitates a progressive geometric reconstruction, evolving from singular points to curves and ultimately to surfaces. Mathematically, this framework ensures the regularity, geometric stability, and topological consistency of the interfaces, thereby providing a rigorously controllable representation for the high-fidelity modeling of complex non-differentiable boundaries.

**Results**

We apply the ILDM to the prediction of 1D and 2D compressible flow fields, as well as the super-resolution (SR) processing of low-resolution data, all of which involve the coexistence of complex non-differentiable interfaces and smooth flow regions. The reconstruction performance is quantitatively evaluated using Mean Squared Error (MSE) and relative error metrics. The 1D benchmarks comprise the Sod shock tube and the peak problems, primarily aimed at benchmarking the ILDM against prevailing physics-informed deep learning models—specifically the R-adaptive DeepONet (RADON) and cPINN—while quantifying the deviations from their respective analytical solutions. The 2D assessments encompass the Riemann problem and the double Mach reflection problem. These tests are designed to evaluate the enhancements provided by the NAF in resolving non-differentiable interfaces, and to examine its compatibility with mainstream deep learning reconstruction models—such as Convolutional Long Short-Term Memory (ConvLSTM), Fourier Neural Operator (FNO), and U-Net—in reconstructing smooth fluid structures. Furthermore, super-resolution experiments are conducted on low-resolution flow fields to assess the capability of the ILDM in recovering fine-scale flow structures from coarse data.

*One-dimensional non-differentiable flow-field reconstruction*

For the first set of experiments, the datasets are generated from the Sod shock-tube problem and the peak problem, which correspond to different evolutions of a one-dimensional Riemann problem under distinct initial conditions. The corresponding formulations are given as follows.

$$\frac{\partial}{\partial t}\begin{pmatrix}\rho \\ \rho u \\ E\end{pmatrix}+\frac{\partial}{\partial x}\begin{pmatrix}\rho u \\ \rho u^2+p \\ u(E+p)\end{pmatrix}=0$$

$$E=\frac{p}{\gamma-1}+\frac{1}{2}\rho u^2$$

For an ideal gas, $\gamma=1.4$ . The Sod problem is the most commonly used one-dimensional benchmark for non-differentiable flow fields, whereas the peak problem exhibits very sharp variations of the physical variables on both sides of the non-differentiable, providing a stringent test of the robustness of the ILDM architecture. For the Sod problem, the initial conditions are set as $(\rho,u,p)_L=(1.0,\quad 0.75,\quad 1.0)$ and $(\rho,u,p)_R=(0.125,\quad 0.0,\quad 0.1)$, with the computational domain $x\in(0,1)$, the initial non-differentiable located at $x_0=0.3$ , and the final time $T=0.2$ . The initial conditions of the peak problem are set as $(\rho,u,p)_L=(0.1261192,8.9047029,782.92899)$ , $(\rho,u,p)_R=(6.591493,2.2654307,3.1544874)$ , with the computational domain $x\in(0,1)$ , the initial non-differentiable initially located at $x_0=0.5$, and the final time $T=0.0039$ . The resulting dataset consists of 500 temporal snapshots of the one-dimensional flow field, with the first 200 time steps used for training and the remaining 300 for testing.

Utilizing these known analytical solutions as test cases, we first quantified the discrepancies between the ILDM predictions and the ground truth, subsequently conducting comparative evaluations of ILDM against RADON and cPINN using the same analytical benchmarks. Figure 2a illustrates the reconstruction errors for the Sod and peak problems, demonstrating that the ILDM consistently maintains a lower error profile, which is significantly smaller than those of the other two models. In Figure 2b, the outputs of the three reconstruction models are compared with the analytical solutions. The results indicate that the ILDM accurately reconstructs the various physical fields in both problems, exhibiting a high degree of fidelity to the analytical solutions in critical regions, such as non-differentiable points and boundaries. While RADON captures the general trend of the analytical solutions, it exhibits pronounced deviations in the localization of non-differentiable points and suffers from numerical oscillations. In contrast, the output of cPINN shows substantial discrepancies from the analytical solutions. Consequently, compared with existing reconstruction models, the ILDM achieves a marked improvement in accuracy when addressing problems characterized by non-differentiable features.

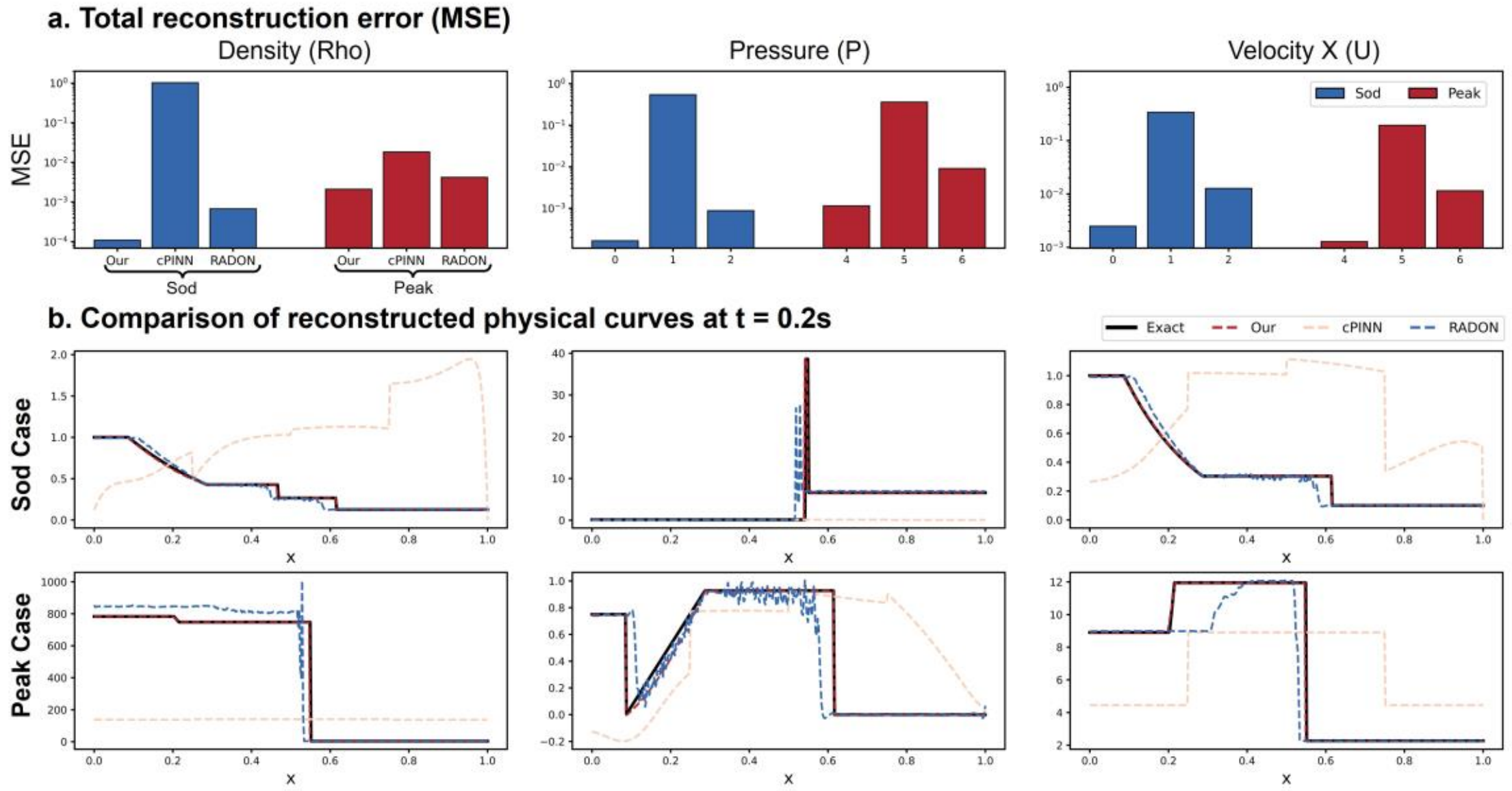

**Fig. 2 | Error statistics and predicted physical profiles of ILDM, cPINN and RADON. a**, Comparison of reconstruction errors for ILDM, cPINN and RADON on the Sod and peak problems. **b**, Comparison of predicted physical profiles obtained by ILDM, cPINN and RADON for the Sod and peak problems.

*Two-dimensional Riemann problems*

The two-dimensional Riemann problem is one of the most classical benchmarks for non-differentiable physical dynamics and is routinely used to assess the performance of numerical algorithms[31]. Its formulation is given as follows.

$$\frac{\partial}{\partial t}\begin{pmatrix}\rho\\ \rho u\\ \rho v\\ E\end{pmatrix}+\frac{\partial}{\partial x}\begin{pmatrix}\rho u\\ \rho u^2+p\\ \rho uv\\ u(E+p)\end{pmatrix}+\frac{\partial}{\partial y}\begin{pmatrix}\rho v\\ \rho uv\\ \rho v^2+p\\ v(E+p)\end{pmatrix}=0$$

$$E = \frac{\rho}{\gamma - 1} + \frac{1}{2}\rho u^2$$

For an ideal gas, $\gamma = 1.4$. The dataset is generated from the two-dimensional Riemann problem with the fluid medium modelled as an ideal gas, on the computational domain $\Omega = [0,1]\times[0,1]$, with the initial conditions specified as follows:

$$\begin{pmatrix}\rho\\u\\v\\p\end{pmatrix}_{x<0.5,y>0.5} = \begin{pmatrix}0.5323\\1.206\\0.0\\0.3\end{pmatrix}, \begin{pmatrix}\rho\\u\\v\\p\end{pmatrix}_{x>0.5,y>0.5} = \begin{pmatrix}1.5\\0.0\\0.0\\1.5\end{pmatrix}$$

$$\begin{pmatrix}\rho\\u\\v\\p\end{pmatrix}_{x<0.5,y<0.5} = \begin{pmatrix}0.138\\1.206\\1.206\\0.029\end{pmatrix}, \begin{pmatrix}\rho\\u\\v\\p\end{pmatrix}_{x>0.5,y<0.5} = \begin{pmatrix}0.5323\\0.0\\1.206\\0.3\end{pmatrix}$$

The dataset contains 200 temporal snapshots of the flow field on a 512*512 grid, with the first 50 time steps used for training and the remaining 150 for testing.

Subsequently, we investigate the specific role of the NAF within the ILDM framework for reconstructing non-differentiable interfaces in complex compressible flows. Figure 3a illustrates the evolution of reconstruction errors over time for models configured both with and without the NAF. The NAF-based model exhibits significantly lower sensitivity to temporal variations, maintaining an error profile that is substantially smaller than that of its counterpart at later time steps. Figure 3b presents high-fidelity simulation snapshots of the 2D Riemann problem alongside the relative errors associated with reconstructions using and omitting the NAF. In regions containing non-differentiable surfaces, the NAF-equipped model yields markedly lower reconstruction errors and also achieves slightly reduced errors in smooth flow regions. These findings demonstrate that the NAF not only ensures the accurate reconstruction of non-differentiable interfaces but also exerts a beneficial influence on the overall reconstruction of smooth fluid structures.

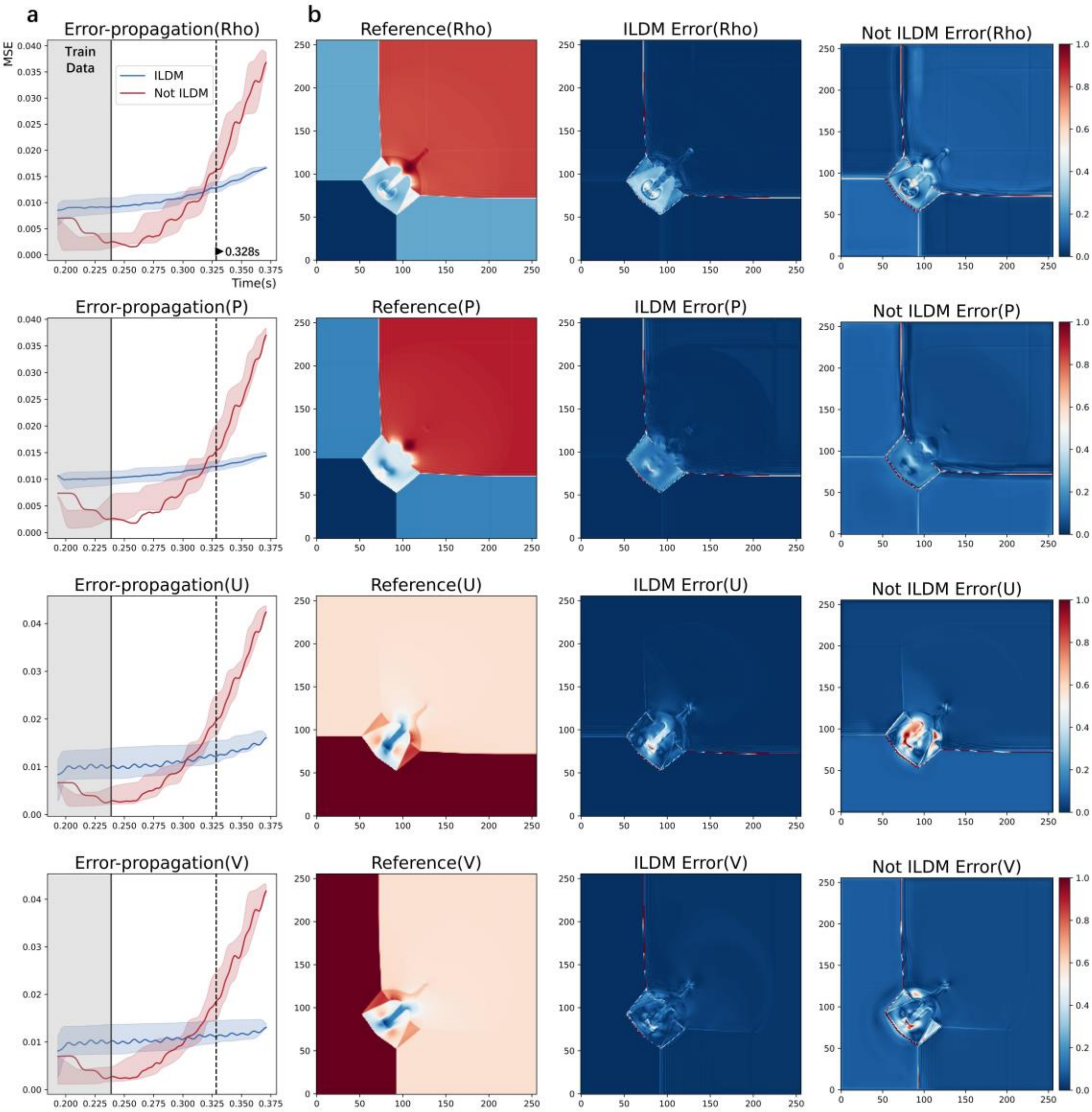

**Fig. 3 | Reconstruction performance with and without NAF when ConvLSTM is used in the SFR interface. a**, Error-propagation curves for models with and without NAF. **b**, Reference compressible physical field and spatial maps of relative error for models with and without NAF.

**Double Mach reflection**

The double Mach reflection problem is a complex flow phenomenon that arises when a strong shock wave impinges obliquely on a ramp[32], and features highly intricate non-differentiable interfaces. We therefore use this benchmark to assess how well mainstream continuous-flow reconstruction models—such as ConvLSTM, FNO and U-Net—can be integrated within ILDM for reconstructing smooth fluid structures, and to identify the most performant reconstruction backbone. The dataset is generated from the double Mach reflection problem with the fluid medium modelled as an ideal gas, on the computational domain $\Omega = [0,4]\times[0,1]$ . The initial condition consists of a strong shock incident at an angle of $60^\circ$ , with the shock initially located at $x_0 = 1/6$, and its incident direction forming a $60^\circ$ angle with the bottom boundary. In the chosen coordinate system, the initial conditions are specified as follows:

$$(\rho,u,v,p) = \begin{cases} (8.0, 8.25\cos(60^\circ), -8.25\sin(60^\circ), 116.5) & , \text{ if } x < x_s(y) \\ (1.4, 0, 0, 1.0) & , \text{ otherwise} \end{cases}$$

where $x_s(y)$ denotes the initial position of the shock.

$$x_s(y) = \frac{1}{6} + \frac{y}{\tan(60^\circ)}$$

The dataset comprises 200 temporal snapshots of the flow field on a (120* 480) grid, with the first 50 time steps used for training and the remaining 150 for testing.

For the Double Mach Reflection problem, we evaluated the reconstruction performance of various deep learning architectures integrated within the ILDM framework. Figure 4a illustrates the temporal evolution of reconstruction errors for ILDM models based on ConvLSTM, FNO, and U-Net. The FNO-based model exhibits substantial reconstruction errors that escalate rapidly over time. In contrast, the models utilizing ConvLSTM and U-Net display comparable error trends, with the ConvLSTM yielding marginally lower overall errors. Noting that both ConvLSTM and U-Net employ an encoder-decoder architecture, we observe that this architectural paradigm demonstrates superior performance in reconstructing smooth fluid regions when coupled with the ILDM. Figure 4b presents the high-fidelity reference snapshots of the Double Mach Reflection alongside the spatial relative error maps for the reconstructions based on ConvLSTM, FNO, and U-Net. The error maps reveal that the error distributions generated by ConvLSTM and U-Net are generally similar and low in magnitude. Conversely, the FNO exhibits striated regions of high error, resulting in significantly elevated reconstruction errors within the smooth fluid zones. These results suggest that encoder-decoder type architectures are more suitable for the reconstruction of smooth fluid structures in this context.

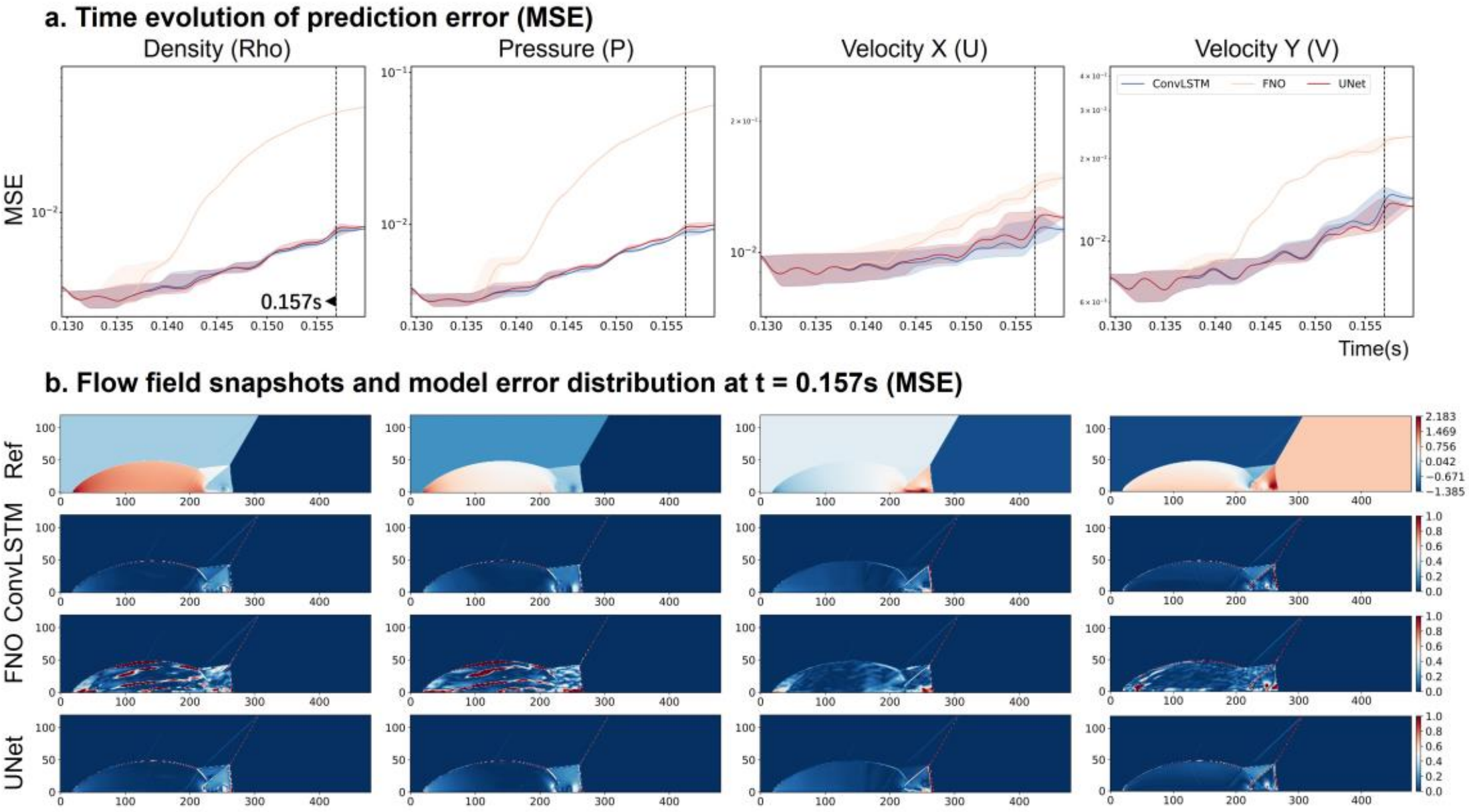

**Fig. 4 | Reconstruction performance of SFR equipped with different smooth fluid reconstruction models under ILDM. a**, Error-propagation curves for different smooth fluid reconstruction models. **b**, Reference compressible physical field and spatial maps of relative error for the different smooth fluid reconstruction models.

*The Propagation Law of Non-differentiable Points*

As the minimal geometric representation within a non-differentiable physical field, the non-differentiable point serves as the foundation for reconstructing non-differentiable interfaces. In this study, we identify non-differentiable points in both the Riemann problem and the double Mach reflection problem, denoting them as DPs. Given the symmetry of the non-differentiable interfaces in the prescribed Riemann flow field with respect to the line Y=X, only the non-differentiable points along the interfaces in the X-direction are labeled. As illustrated in Figure 5, the DPs predicted by the NAF exhibit a high degree of consistency in propagation characteristics with those identified in the reference field. This predictive accuracy ensures the positional precision of the subsequent non-differentiable interface reconstruction.

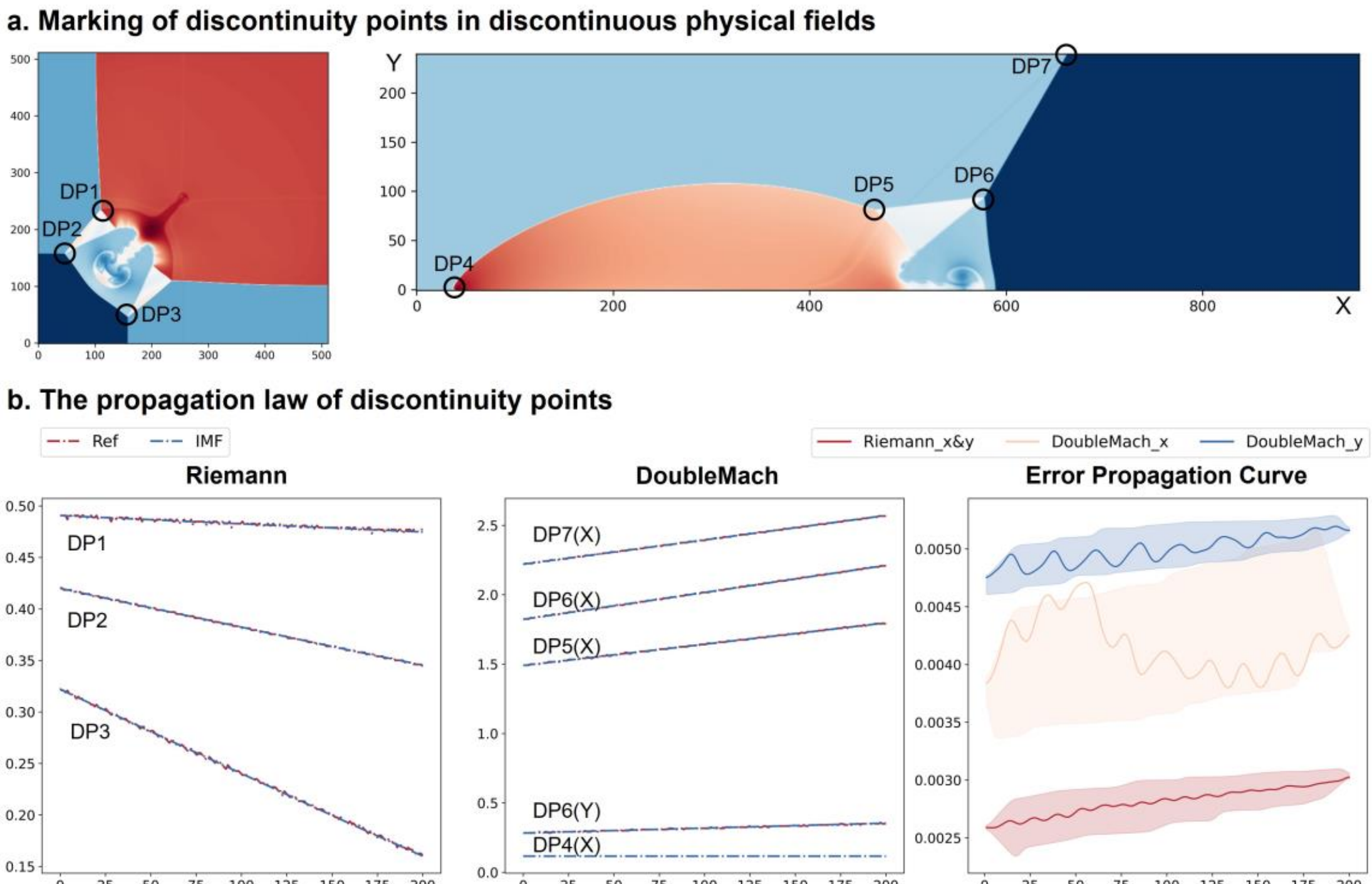

**Fig. 5 | Propagation characteristics of non-differentiable points. a**, Identification of non-differentiable points within the compressible physical field. Error-propagation curves for different smooth fluid reconstruction models. **b**, Comparison between the ILDM predicted non-differentiable points and the labeled points in the reference compressible physical field, including the Mean Squared Error (MSE).

*Super-resolution*

Training datasets for our model typically require high-resolution non-differentiable fields, whose generation is computationally expensive, whereas super-resolution techniques can substantially reduce the cost of data acquisition. Noting that, within ILDM, the accuracy of non-differentiable-interface reconstruction is not directly tied to the resolution of the input flow field, we anticipate that ILDM may exhibit strong performance on super-resolution tasks. To examine this, we perform spatio-temporal subsampling of the two-dimensional Riemann and double Mach reflection problems, with a temporal sampling interval of 5 time steps. For the two-dimensional Riemann problem, the spatially downsampled resolutions are 32*32, 64*64, 128*128 and 256*256; for the double Mach reflection problem, the corresponding resolutions are 15*60, 30*120, 60*240 and 120*480.

We next examine the super-resolution performance of ILDM under different low-resolution inputs. In Figure 6a, we show the reconstruction errors of ILDM for the double Mach reflection super-resolution task, and in Figure 6b the corresponding errors for the Riemann problem. The curves indicate that ILDM maintains a stable super-resolution accuracy across different problems and resolutions, while keeping the reconstruction error at a low level. Figure 6c displays input snapshots of the Riemann problem at different resolutions together with their reconstructed compressible physical fields. The error maps reveal that, although some reconstruction discrepancies arise in the smooth fluid regions at different upsampling levels, the reconstructed non-differentiable regions remain highly consistent. Taken together, both the error curves and the reconstruction contours demonstrate that the NAF and SFR components of ILDM are largely insensitive to input resolution in super-resolution tasks, yielding high stability and consistency.

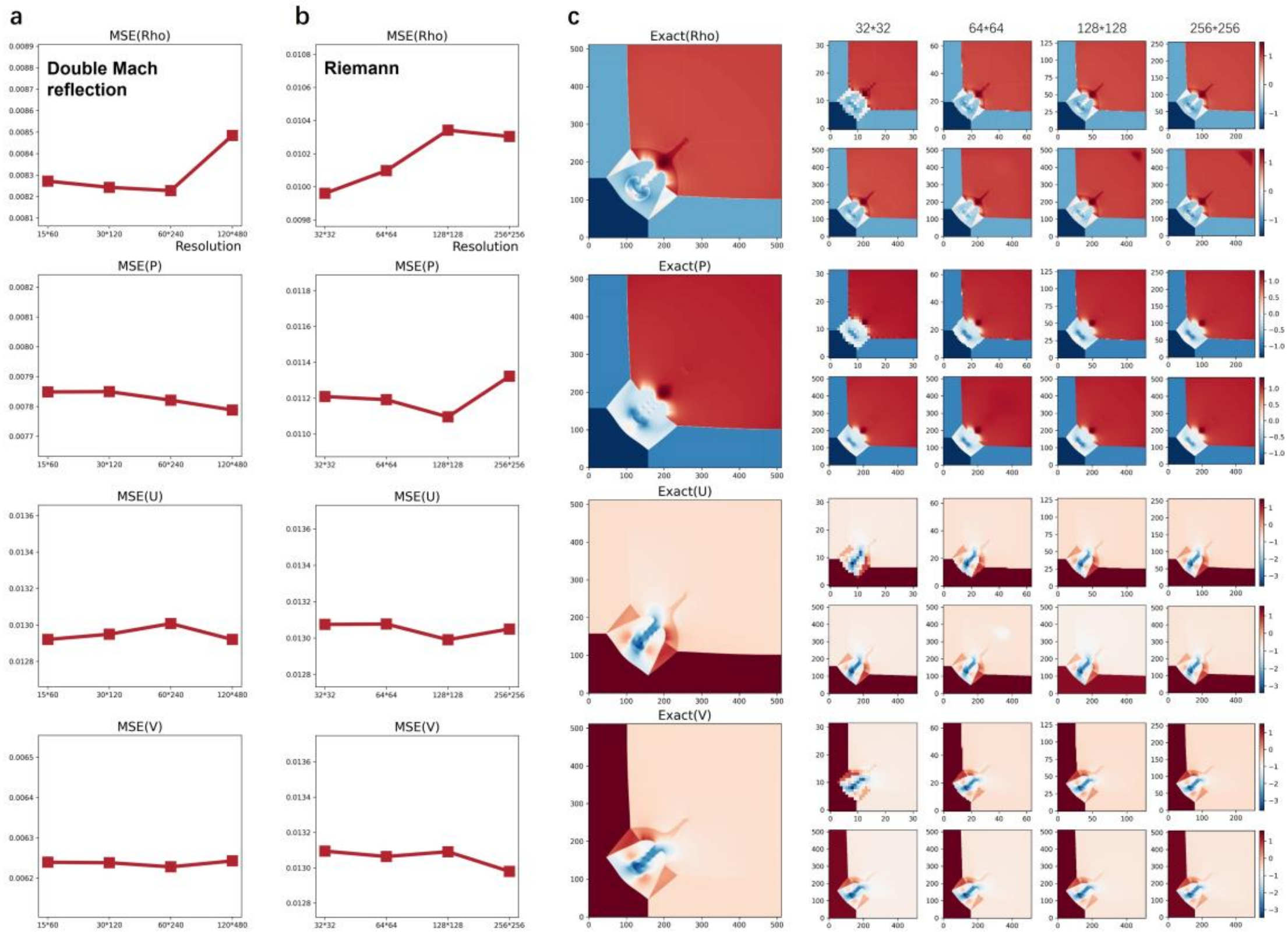

**Fig. 6 | Super-resolution errors and reconstructed compressible physical field snapshots at different resolutions.** **a**, Comparison of super-resolution reconstruction errors at different input resolutions for the double Mach reflection problem. **b**, Comparison of super-resolution reconstruction errors at different input resolutions for

the Riemann problem. **c**, Reference high-fidelity snapshot of the Riemann problem, together with input flow-field snapshots at different resolutions and their corresponding reconstructed fields.

**Discussion**

In this work, we present the Inverse Low-Dimensional Manifold reconstruction framework, a unified framework for the reconstruction of compressible flow fields leveraging the NAF and SFR. The central premise lies in utilizing the NAF with explicit parameters to approximate the gradient characteristics inherent to non-differentiable interfaces. Building upon this, the ILDM mapping is employed to recover spatial interface features by progressively performing dimensionality lifting of geometric entities, evolving from localized non-differentiable features to global interfaces. This modeling strategy enables the stable representation of intricate non-differentiable structures while preserving physical consistency, thereby offering a novel perspective on spatiotemporal reconstruction challenges in compressible physical fields.

Compared with conventional architectures such as cPINN and R-Adaptive DeepONet, the principal advantage of the ILDM lies in its ability to reconstruct non-differentiable interfaces directly according to their propagation laws, thereby obviating the need for manual specification of subdomain boundaries. By explicitly characterizing the localization and gradient features of singular points via the NAF, the framework effectively circumvents numerical oscillations typically induced by gradient singularities. Furthermore, the progressive, dimensionality-lifting reconstruction paradigm inherent in the ILDM framework establishes a unified mathematical pathway for the precise modeling of intricate interface geometries. The ILDM framework is also highly extensible, allowing for the seamless integration of various smooth-flow reconstruction modules—such as ConvLSTM and U-Net—to enhance its representational capacity for complex smooth fluids. Numerical results demonstrate that the proposed framework achieves reconstruction accuracy significantly superior to mainstream benchmarks across 1D and 2D non-differentiable flows, including the Sod shock tube, peak problems, 2D Riemann problems, and double Mach reflections. Notably, it maintains robust super-resolution performance even when trained on low-resolution inputs.

The architecture of the ILDM unifies the reconstruction of non-differentiable interfaces and smooth flow regions, thereby establishing an equilibrium between physical consistency and numerical generalization. Nevertheless, within the current framework, the gradient scaling factors of the NAF necessitate pre-configuration based on specific initial conditions. In future research, these parameters could be integrated into the loss function and optimized concurrently with the network, facilitating a fully data-driven calibration of the structural functions. Furthermore, while the current smooth-flow module is primarily tailored for deep learning-based architectures, extending this interface to incorporate traditional numerical solvers would significantly broaden the framework’s applicability and potentially enhance its overall reconstruction precision.

In summary, the ILDM framework proposed in this study provides a novel theoretical paradigm and a robust computational tool for the high-fidelity reconstruction of compressible physical fields. The methodology not only yields superior performance in canonical fluid dynamics benchmarks but also demonstrates significant potential for extension to multi-physics coupled systems, such as detonation and phase transitions. Future endeavors will prioritize enhancing the framework's computational scalability, aiming to facilitate more universal and physically consistent reconstruction and prediction for complex nonlinear systems.

## Appendix A. Validation of the method

*Design rationale and validation of NAF*

The design of $F_{\varphi,\phi}(x;v^i,x^i)$ is inspired by the characterization of non-differentiable jumps using the Heaviside function, which represents non-differentiable structures in physical fields in their simplest possible form.

$$H(x)=\begin{cases}0, & x\le 0,\\ 1, & x>0\end{cases}$$

However, it cannot capture complex non-differentiable patterns in realistic physical fields. We therefore introduce a function $F_{\varphi,\phi}(x;v^i,x^i)$ capable of representing intricate local transition profiles. The structure of $F_{\varphi,\phi}(x;v^i,x^i)$ is constructed from sigmoid functions equipped with scale factors:

$$\sigma_{\varphi}(x)=\frac{1}{1+e^{-x*\varphi}}$$

where $\varphi$ denotes a tunable scale factor; when $\varphi\to\infty$, $\sigma_{\varphi}(x)$ mathematically converges to the Heaviside function $H(x)$:

$$\lim_{\varphi\to 0}\sigma_{\varphi}(x)=H(x)=\begin{cases}0, & x\le 0,\\ 1, & x>0\end{cases}$$

When constructing

$$S(x)=\sigma_{\varphi}(x)\sigma_{\phi}(-x),$$

When $\varphi,\phi\to\infty$, the transition becomes infinitely steep, and the product ensures that the gradient of the function is non-zero only in the vicinity of the interface and approaches zero away from it, as illustrated by the gradient profile in Fig. 1a. Accordingly, $\sigma_{\varphi}(x)\sigma_{\phi}(-x)$ serves as an approximate Heaviside indicator function with a tunable gradient. The function $F_{\varphi,\phi}(x;v^{i},x^{i})$ further augments $\sigma_{\varphi}(x)\sigma_{\phi}(-x)$ by introducing a scaling coefficient $v^{i}$ and a position variable $x^{i}$:

$$v^{i}*\frac{1}{1+e^{(x-x^{i})*\varphi}}*\frac{1}{1+e^{-(x-x^{i})*\phi}}$$

which does not alter its ability to represent the underlying non-differentiable.

*Validation of the regularity and topological consistency of ILDM*

Using ILDM, we achieve a progressive reconstruction of the geometric features of non-differentiable interfaces. Here, we employ Lipschitz continuity to assess whether ILDM guarantees interface regularity, geometric stability and topological consistency.

First, let $u:\Omega\subset\mathbb{R}^{n}\to\mathbb{R}$ denote a piecewise smooth physical field. Its non-differentiable set $\Gamma=\{x\in\Omega:u(x^{-})\neq u(x^{+})\}$ can, in the sense of geometric measure theory, be regarded as a $(n-1)$-dimensional bounded Lipschitz manifold, i.e. $\Gamma\in\mathcal{M}_{n-1}^{\mathrm{Lip}}(\Omega)$.

The non-differentiable curves and points on this interface then constitute its 1D and 0D Lipschitz submanifolds, respectively:

$\Gamma_{1}\subset\Gamma,\quad\Gamma_{0}\subset\Gamma_{1}.$

We define, in the parametric Non-differentiable Structural Function, $F_{\varphi,\phi}:\Omega\to\mathbb{R}$, with respect to any finite parameter vector $\varphi,\phi$, and denote the corresponding mapping by $F_{\varphi,\phi}\in\mathrm{Lip}(\Omega)$, ; then $F_{\varphi,\phi}$ is globally Lipschitz continuous and satisfies

$\|F_{\varphi,\phi}(x)-F_{\varphi,\phi}(y)\|\le L_{\varphi,\phi}\|x-y\|,\quad\forall x,y\in\Omega,$

where $L_{\varphi,\phi}$ is the Lipschitz constant controlled by the sigmoid scale factors $\varphi,\phi$.
As the scale factors approach their limiting values (for example, in the case of a sharpened sigmoid), $F_{\varphi,\phi}$ converges in the sense of distributions to a Heaviside-type interface indicator function $\chi_\Gamma(x),$, while preserving a controllable Lipschitz upper bound.

For any $n \in \{0,1\}$, we define the level-set mapping $\Phi_k : \mathrm{Lip}(\Omega) \to \mathcal{M}_{n+1}^{\mathrm{Lip}}(\Omega)$ as
$\mathcal{M}_{n+1} = \Phi_n(F_{\varphi,\phi}) = \{x \in \Omega : F_{\varphi,\phi}(x) = c_n\},$
where $c_n$ is a constant threshold.

Since $F_{\varphi,\phi}$ is a Lipschitz continuous function, its level sets are Lipschitz manifolds in the sense of geometric measure theory (satisfying the implicit-function condition almost everywhere), and hence $\mathcal{M}_{n+1} \in \mathcal{M}_{n+1}^{\mathrm{Lip}}(\Omega).$. This implies that, provided the Lipschitz regularity of (d) is ensured, the interface curves and surfaces induced by $F_{\varphi,\phi}$ possess good geometric regularity (for example, bounded curvature, finite measure and the absence of sharp singularities).

Through the mapping chain $\mathcal{M}_0 \to \mathcal{M}_1 \to \mathcal{M}_2$, we thus achieve a hierarchical reconstruction from point-like discontinuities to curves and then to full interfaces. Mathematically, this corresponds to a family of embedding maps satisfying Lipschitz regularity, $\Phi_0 : \mathcal{M}_0 \to \mathcal{M}_1, \quad \Phi_1 : \mathcal{M}_1 \to \mathcal{M}_2,$
which ensure that the geometric objects at each level retain controlled interface smoothness and topological consistency.